\documentclass[reprint, amsmath,amssymb,aps,prd,]{revtex4-1}
\usepackage{subfigure}
\usepackage{graphicx}
\usepackage{dcolumn}
\usepackage{bm}
\usepackage{xcolor}
\definecolor{aa}{RGB}{0,0,139}

\newcommand{\B}{\mathcal{B}}
\usepackage[colorlinks,urlcolor=aa,linkcolor=blue,anchorcolor=aa,citecolor=aa]{hyperref}
\usepackage[mathlines]{lineno}
\usepackage{multirow}
\usepackage{enumerate}
\usepackage{amsmath}
\begin{document}

\preprint{APS/123-QED}

\title{{\bf \boldmath Measurements of the branching fractions of
$\psi(3686)\rightarrow\bar{\Sigma}^{0}\Lambda+c.c.$ and $\chi_{cJ (J
= 0,1,2)} \rightarrow \Lambda \bar{\Lambda}$}}

\author{M.~Ablikim$^{1}$, M.~N.~Achasov$^{10,c}$, P.~Adlarson$^{67}$, S. ~Ahmed$^{15}$, M.~Albrecht$^{4}$, R.~Aliberti$^{28}$, A.~Amoroso$^{66A,66C}$, M.~R.~An$^{32}$, Q.~An$^{63,49}$, X.~H.~Bai$^{57}$
% ...
\\
\vspace{0.2cm}
(BESIII Collaboration)\\
\vspace{0.2cm} {\it
$^{1}$ Institute of High Energy Physics, Beijing 100049, People's Republic of China\\
% ...
$^{4}$ Bochum Ruhr-University, D-44780 Bochum, Germany\\
% ...
$^{10}$ G.I. Budker Institute of Nuclear Physics SB RAS (BINP), Novosibirsk 630090, Russia\\
% ...
$^{15}$ Helmholtz Institute Mainz, Staudinger Weg 18, D-55099 Mainz, Germany\\
% ...
$^{28}$ Johannes Gutenberg University of Mainz, Johann-Joachim-Becher-Weg 45, D-55099 Mainz, Germany\\
% ...
$^{32}$ Liaoning Normal University, Dalian 116029, People's Republic of China\\
% ...
$^{49}$ State Key Laboratory of Particle Detection and Electronics, Beijing 100049, Hefei 230026, People's Republic of China\\
% ...
$^{57}$ University of Jinan, Jinan 250022, People's Republic of China\\
% ...
$^{63}$ University of Science and Technology of China, Hefei 230026, People's Republic of China\\
% ...
$^{66}$ University of Turin and INFN, (A)University of Turin, I-10125, Turin, Italy; (B)University of Eastern Piedmont, I-15121, Alessandria, Italy; (C)INFN, I-10125, Turin, Italy\\
$^{67}$ Uppsala University, Box 516, SE-75120 Uppsala, Sweden\\
% ...
$^{c}$ Also at the Novosibirsk State University, Novosibirsk, 630090, Russia\\
% ...
}}

\vspace{0.4cm}
\date{\today}
\begin{abstract}
   Based on $4.481\times10^8$ $\psi(3686)$ events collected with the
   BESIII detector at BEPCII, the branching fraction of the isospin
   violating decay $\psi(3686)\rightarrow\bar{\Sigma}^{0}\Lambda+c.c.$
   is measured to be $(1.60 \pm 0.31 \pm 0.13~\pm~0.58) \times
   10^{-6}$, where the first uncertainty is statistical, the second is
   systematic, and the third is the uncertainty arising from
   interference with the continuum. This result is significantly
   smaller than the measurement based on CLEO-c data sets. The decays
   $\chi_{cJ} \rightarrow\Lambda\bar{\Lambda} $ are measured via
   $\psi(3686)\rightarrow\gamma\chi_{cJ}$, and the branching fractions
   are determined to be
   $\B\left(\chi_{c0}\rightarrow\Lambda\bar{\Lambda}\right)=(3.64 \pm
   0.10 \pm 0.10 \pm 0.07)\times 10^{-4}$,
   $\B\left(\chi_{c1}\rightarrow\Lambda\bar{\Lambda}\right)=(1.31\pm0.06
   \pm 0.06 \pm0.03)\times 10^{-4}$,
   $\B\left(\chi_{c2}\rightarrow\Lambda\bar{\Lambda}\right)=(1.91\pm0.08
   \pm 0.17 \pm0.04)\times 10^{-4}$, where the third uncertainties are
   systematic due to the $\psi(3686) \rightarrow
   \gamma \chi_{c J}$ branching fractions.
\end{abstract}

\pacs{Valid PACS appear here}
\maketitle

\section{\label{sec:introduction}Introduction}

Experimental studies of charmonium decays are essential for
understanding the structures and decay mechanisms of charmonium states.
These measurements enable tests of non-perturbative Quantum Chromodynamics (QCD) models. Further, charmonium decays to baryon pairs
provide a novel method to explore the properties of
baryons~\cite{a3,polarization}. In recent years, there have been
searches for missing decays and measurements of angular distributions
and polarizations of many $J/\psi$ and $\psi(3686)$ two-body decays
to baryon and anti-baryon final states with much improved precision by
the CLEO, BESII, and BESIII
collaborations~\cite{CLEO-measurement,BES-measurement1,BES-measurement2,BES-measurement3,BESIII-measurement1,BESIII-measurement2,BESIII-measurement3,1803.02039,1907.13041,1911.06669,2004.07701,2007.03679}.
In addition, these measurements also provide the possibility to
determine the relative phase between strong and electro-magnetic
amplitudes~\cite{a4}.

In spite of the significant improvements achieved on
 $J/\psi$ and $\psi(3686)$  decays into baryon pairs, information
on isospin symmetry breaking decays is still limited due to their low
decay rates. Recently, a measurement of the
isospin violating decay
$\psi(3686)\rightarrow\bar{\Sigma}^{0}\Lambda+c.c.$ is reported and obtained a
branching fraction of
$\B\left(\psi(3686)\rightarrow\bar{\Sigma}^{0}\Lambda+c.c.\right)=(12.3~\pm~2.4)\times10^{-6}$
based on $2.45\times10^7~\psi(3686)$ decay events collected with CLEO-c detector~\cite{a2}. This result is much larger than theoretical predictions~\cite{a4}, and a specific mechanism is proposed by the authors of~\cite{Ferroli:2020xnv} to explain its abnormal largeness.
In our analysis, we use a sample of
$4.481\times10^8$ $\psi(3686)$ events collected at BESIII to measure
the branching fraction of $\psi(3686) \rightarrow \bar{\Sigma}^{0}
\Lambda +c . c .$, with $\bar{\Sigma}^{0}\rightarrow \gamma
\bar{\Lambda}$,~$\bar{\Lambda}~(\Lambda) \rightarrow \bar{p} \pi^{+}
(p \pi^{-})$.

With the same final states, we can also study the
$\Lambda\bar{\Lambda}$ pair decay from the P-wave charmonium
$\chi_{cJ}$ states, which are produced via a radiative transition from
the $\psi(3686)$. Using $ 1.068 \times 10^{8}~\psi(3686)$ events
collected in $2009,$ BESIII previously reported the branching-fraction
measurements of $\B \left(\chi_{c 0} \rightarrow \Lambda
\bar{\Lambda}\right)=(33.3 \pm 2.0 \pm 2.6) \times 10^{-5}$
$\B\left(\chi_{c 1} \rightarrow \Lambda \bar{\Lambda}\right)=(12.2 \pm
1.1 \pm 1.1) \times 10^{-5},$ and $\B \left(\chi_{c 2}
\rightarrow\right.$ $\Lambda \bar{\Lambda})=(20.8 \pm 1.6 \pm 2.3)
\times 10^{-5}$~\cite{Beschicj}.

\section{\label{detector}Detector and Monte Carlo simulation}

The BESIII detector~\cite{Ablikim:2009aa} records symmetric $e^+e^-$ collisions
provided by the BEPCII storage ring~\cite{Yu:IPAC2016-TUYA01}, which operates with a peak luminosity of $1\times10^{33}$~cm$^{-2}$s$^{-1}$
in the center-of-mass energy range from 2.0 to 4.7~GeV.
BESIII has collected large data samples in this energy region~\cite{Ablikim:2019hff}. The cylindrical core of the BESIII detector covers 93\% of the full solid angle and consists of a helium-based multilayer drift chamber~(MDC), a plastic scintillator time-of-flight
system~(TOF), and a CsI(Tl) electromagnetic calorimeter~(EMC),
which are all enclosed in a superconducting solenoidal magnet
providing a 1.0~T magnetic field. The solenoid is supported by an
octagonal flux-return yoke with resistive plate counter muon
identification modules interleaved with steel.
The acceptance of charged particles and photons is 93\% of the 4$\pi$ solid angle.
The charged-particle momentum resolution at $1~{\rm GeV}/c$ is
$0.5\%$, and the $dE/dx$ resolution is $6\%$ for electrons
from Bhabha scattering. The EMC measures photon energies with a
resolution of $2.5\%$ ($5\%$) at $1$~GeV in the barrel (end cap)
region. The time resolution in the TOF barrel region is 68~ps, while
that in the end cap region is 110~ps.

Simulated Monte Carlo (MC) samples produced with {\sc
  geant4}-based~\cite{geant4} software, which includes the geometric
description of the BESIII detector and the detector response, are used
to determine detection efficiencies, estimate background
contributions, and study systematic uncertainties. The simulation models
the beam energy spread and initial state radiation (ISR) in the
$e^+e^-$ annihilations with the generator {\sc
  kkmc}~\cite{ref:kkmc}. The inclusive MC sample simulates every possible process, who includes the production of the $\psi(3686)$ resonance, the ISR production of the
$J/\psi$, and the continuum processes incorporated in {\sc
  kkmc}~\cite{ref:kkmc}. The known decay modes are modeled with {\sc
  evtgen}~\cite{ref:evtgen} using branching fractions taken from the
Particle Data Group~\cite{PDG}, and the remaining unknown charmonium
decays are modeled with {\sc lundcharm}~\cite{ref:lundcharm}. Final
state radiation~(FSR) from charged particles is
incorporated using {\sc photos}~\cite{photos}. For the
signal processes, we use MC samples of
$\psi(3686)\rightarrow\bar{\Sigma}^{0}\Lambda + c.c.$ decays generated
with uniform phase space (PHSP), while
$\psi(3686)\rightarrow\gamma\chi_{cJ}$ decays are generated according
to helicity amplitudes~\cite{angleUncer} and $\chi_{cJ} \to
\Lambda\bar{\Lambda}$ with PHSP.

\section{\boldmath $\psi(3686) \rightarrow \bar{\Sigma}^{0} \Lambda +c . c .$}
\subsection{\label{3.2}Event selection}
Since the final state of interest is $\gamma p \bar{p} \pi^{+}
\pi^{-},$ we require each $\psi(3686)$ candidate to contain four
charged tracks with zero net charge, and at least one photon. Each charged
track, detected in the MDC is required to satisfy
$|\rm{cos\theta}|<0.93$, where $\theta$ is defined with respect to the
$z$-axis, which is the symmetry axis of the MDC. The distance of
closest approach to the interaction point (IP) must be less than
30\,cm along the $z$-axis, and less than 10\,cm in the transverse
plane. Pions and protons are identified by the magnitude of their
momentum, and charged tracks with momentum larger than 0.7
$\mathrm{GeV } /c$ in the lab frame are identified as protons. Other
tracks are identified as pions. An isolated cluster in the EMC is
considered to be a photon if the following requirements are satisfied:
1) the deposited energy of each shower must be more than 25~MeV in the
barrel region ($|\!\cos \theta|< 0.80$) and more than 50~MeV in the
end cap region ($0.86 <|\!\cos \theta|< 0.92$); 2) to suppress
electronic noise and showers unrelated to the event, the difference
between the EMC time and the event start time is required to be within
(0, 700)\,ns; 3) to exclude showers that originate from charged
tracks, the angle between the position of each shower in the EMC and
the closest extrapolated charged track of $p, \pi^{+},$ or $\pi^{-}$
must be greater than 10 degrees, and greater than 20 degrees for the
$\bar{p}$ track.

The $\Lambda(\bar{\Lambda})$ candidate is reconstructed with any
$p\pi^-$~$(\bar{p}\pi^+)$~combination satisfying a secondary vertex
fit~\cite{secondvertex}. The secondary vertex fit is required to be
successful, but no additional requirements are placed on the fit
$\chi^{2}$. To improve the momentum and energy resolution and to
reduce background contributions, a six-constraint (6C) kinematic fit
is applied to the event candidates with constraints on the total
four-momentum and the invariant masses of the $\Lambda$
and $\bar{\Lambda}$ candidates. The $\chi_{6 C}^{2}$ of the kinematic
fit is required to be less than 25.

To further suppress background, we require: 1) the $\chi_{6 C}^{2}$
for the $\gamma p \bar{p} \pi^+ \pi^-$ hypothesis is less than those
for any $\gamma \gamma p \bar{p} \pi^+ \pi^-$ or $ p \bar{p} \pi^+
\pi^-$ hypothesis: $\chi_{\gamma p\bar{p} \pi^{+}
  \pi^{-}}^{2}<\chi_{\gamma\gamma p\bar{p}\pi^{+}\pi^{-}}^{2}$,
$\chi_{\gamma p\bar{p}\pi^{+}\pi^{-}}^{2}<\chi_{
  p\bar{p}\pi^{+}\pi^{-}}^{2} $; 2) the $\Lambda$($\bar{\Lambda})$
lifetime must satisfy $L_{\Lambda(\bar{\Lambda})} / \sigma>2$
where $L$ and $\sigma$ are the decay length and its
uncertainty obtained from the secondary vertex fit; 3) the invariant
mass of the $\Lambda\bar{\Lambda}$, $M_{\Lambda \bar{\Lambda}}$ is
required to be greater than 3.48 $\mathrm{GeV} /c^{2}$ in order to
suppress the $\psi(3686) \rightarrow \gamma \chi_{c 0},~\chi_{c 0}
\rightarrow \Lambda \bar{\Lambda}$ background; 4) $\rm M_{\gamma
  \Lambda}>1.15$ $\mathrm{GeV} /c^{2}$ and $\rm M_{\gamma
  \bar{\Lambda}}>1.15$ $\mathrm{GeV} /c^{2}$ are required to suppress
background from $\psi(3686) \rightarrow \Lambda \bar{\Lambda}$.

\begin{figure*}[htb]
  \centering
  \mbox{
  \subfigure{\includegraphics[width=0.45\textwidth]{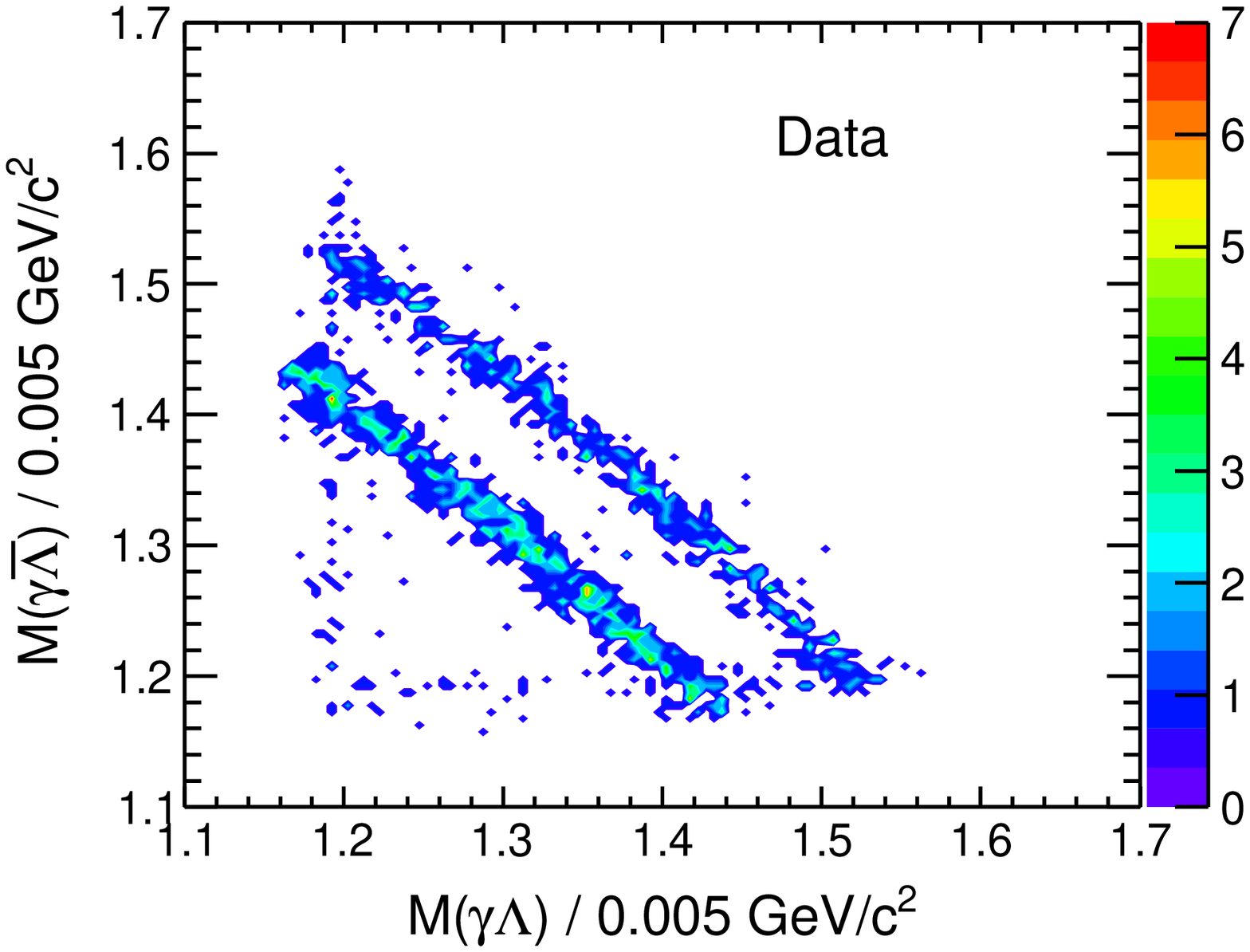}}\put(-180,140){\bf  ~(a)}
  \subfigure{\includegraphics[width=0.45\textwidth]{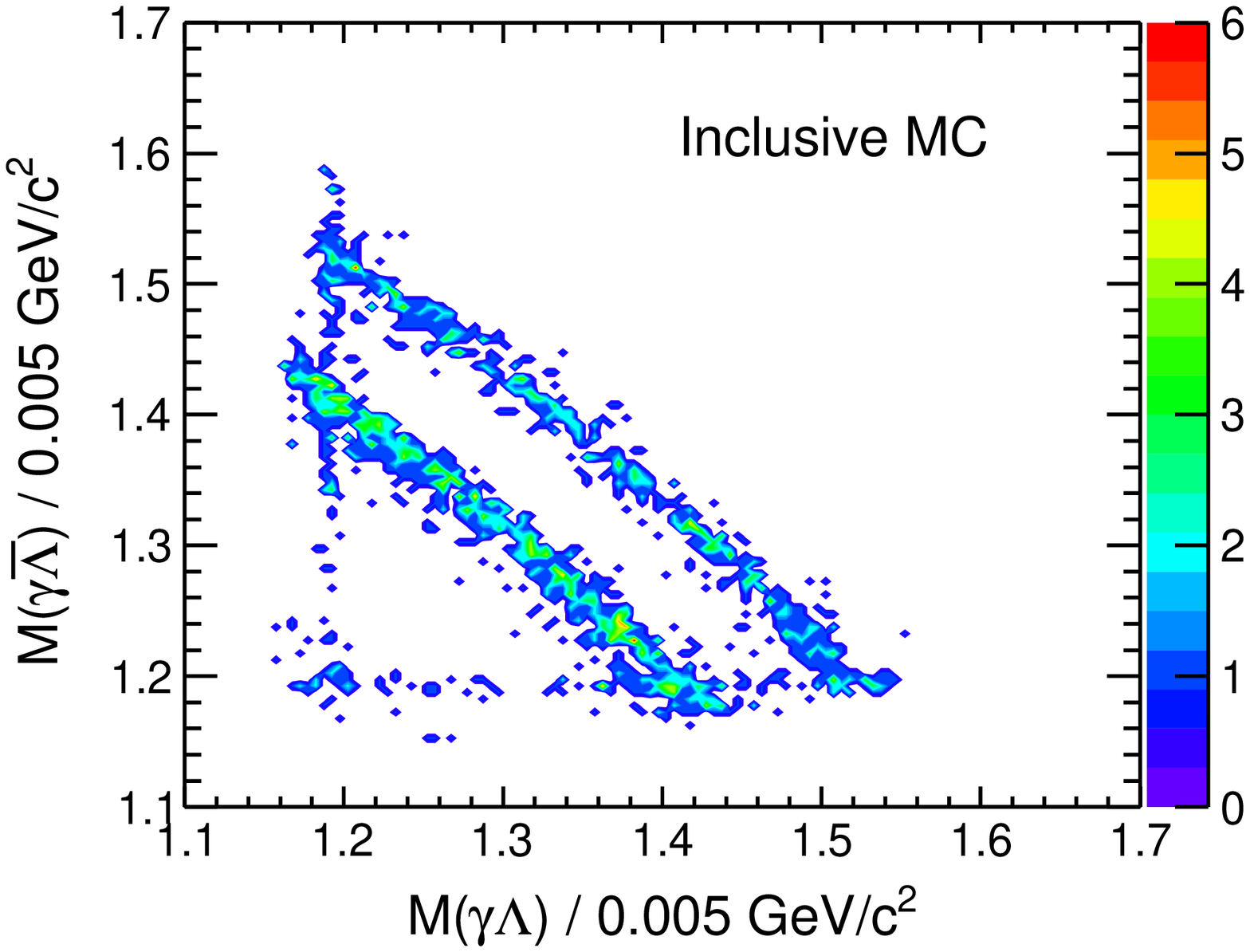}}\put(-180,140){\bf  ~(b)}
  }
  \centering
  \mbox{
  \subfigure{\includegraphics[width=0.45\textwidth]{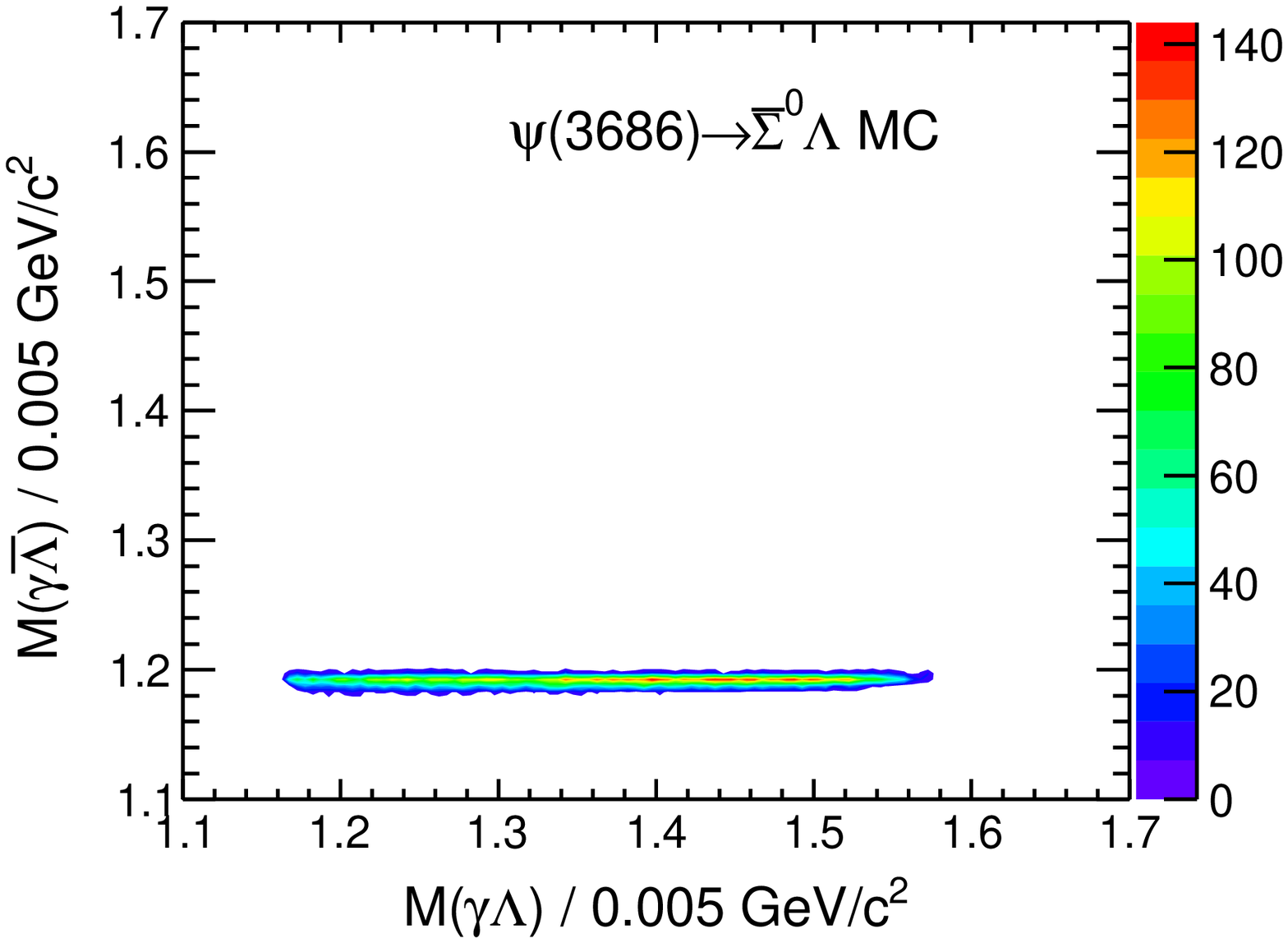}}\put(-180,140){\bf  ~(c)}
  \subfigure{\includegraphics[width=0.45\textwidth]{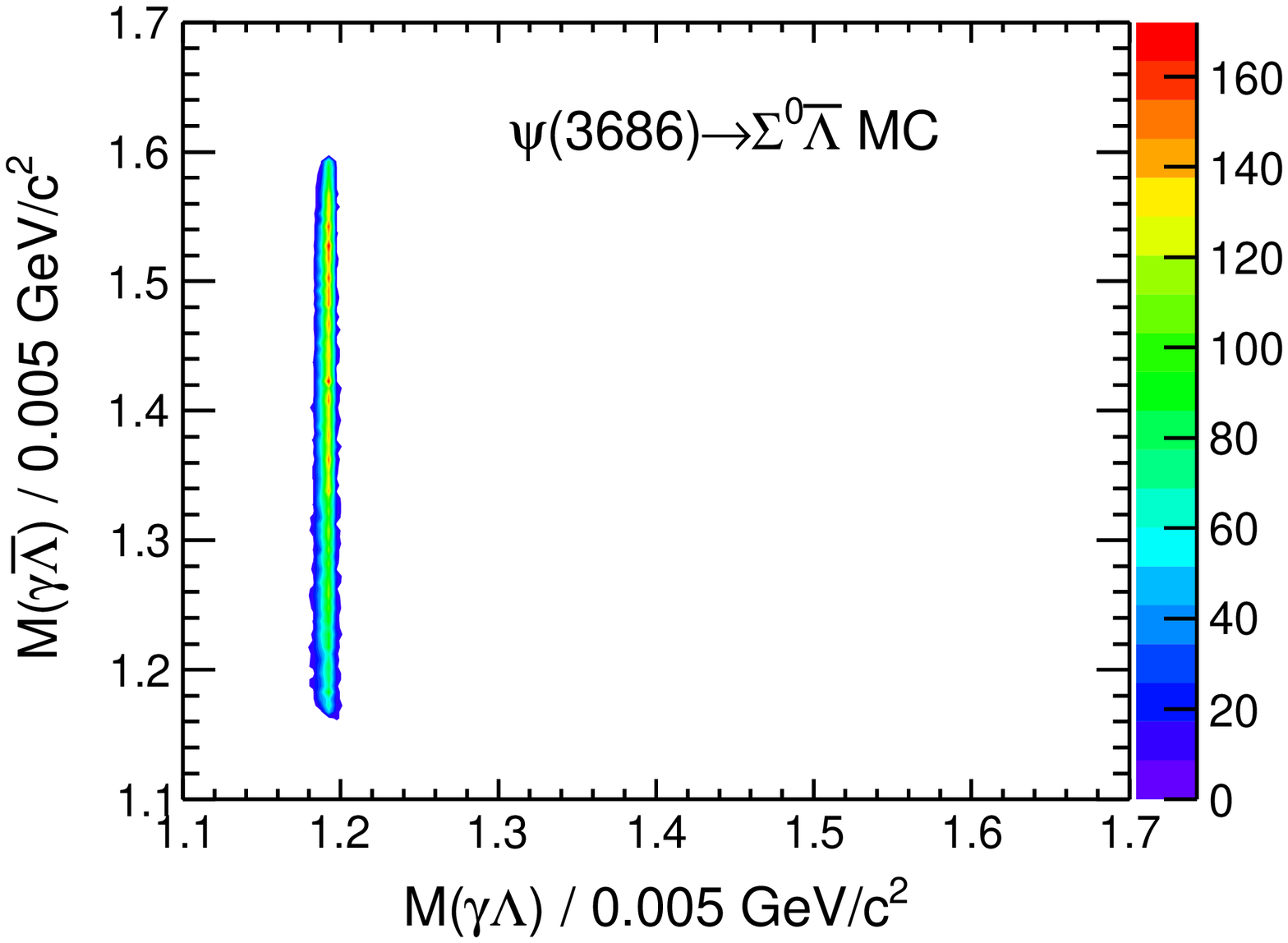}}\put(-180,140){\bf  ~(d)}
  }
  \caption{ Scatter distributions of $\rm
    M_{\gamma\bar{\Lambda}}$ versus $\rm M_{\gamma\Lambda}$ at the
    $\psi(3686)$ resonance. (a), (b), (c) and (d) are distributions from data,
    inclusive MC samples, and signal MC samples of
    $\psi(3686)\rightarrow\bar{\Sigma}^{0}\Lambda$ and
    $\psi(3686)\rightarrow\Sigma^{0}\bar{\Lambda}$ decays,
    respectively.}
 \label{bkgso}
\end{figure*}

After imposing the above requirements, Fig.~\ref{bkgso} shows the
scatter plots of $\rm M_{\gamma\bar{\Lambda}}$ versus $\rm
M_{\gamma\Lambda}$ for data, inclusive MC samples, and signal MC samples
of $\psi(3686)\rightarrow\bar{\Sigma}^{0}\Lambda$ and
$\psi(3686)\rightarrow\Sigma^{0}\bar{\Lambda}$ processes. The $\psi(3686)
\rightarrow \bar{\Sigma}^{0}\Lambda+c.c.$ signals are clearly visible in Fig.~\ref{bkgso}~(a). The two
sloped bands are backgrounds from $\psi(3686)\rightarrow \gamma
\chi_{c1,2}, \chi_{c1,2}\rightarrow \Lambda \bar{\Lambda}$, which are
well simulated by the inclusive MC samples as shown in
Fig.~\ref{bkgso}~(b).  The inclusive MC indicates that the only
peaking background is from $\psi(3686)\rightarrow \bar{\Sigma}^{0}
\Sigma^{0}$. Figures~\ref{bkgso}~(c) and (d) are the signal shapes of
$\psi(3686)\rightarrow\bar{\Sigma}^{0}\Lambda$ and
$\psi(3686)\rightarrow\Sigma^{0}\bar{\Lambda}$, which are vertical and
horizonal bands in the scatter plot distributions, respectively.

\subsection{Signal yields and branching fraction calculation}
\begin{figure*}[htb]
  \centering
  \subfigure{\includegraphics[width=0.48\textwidth]{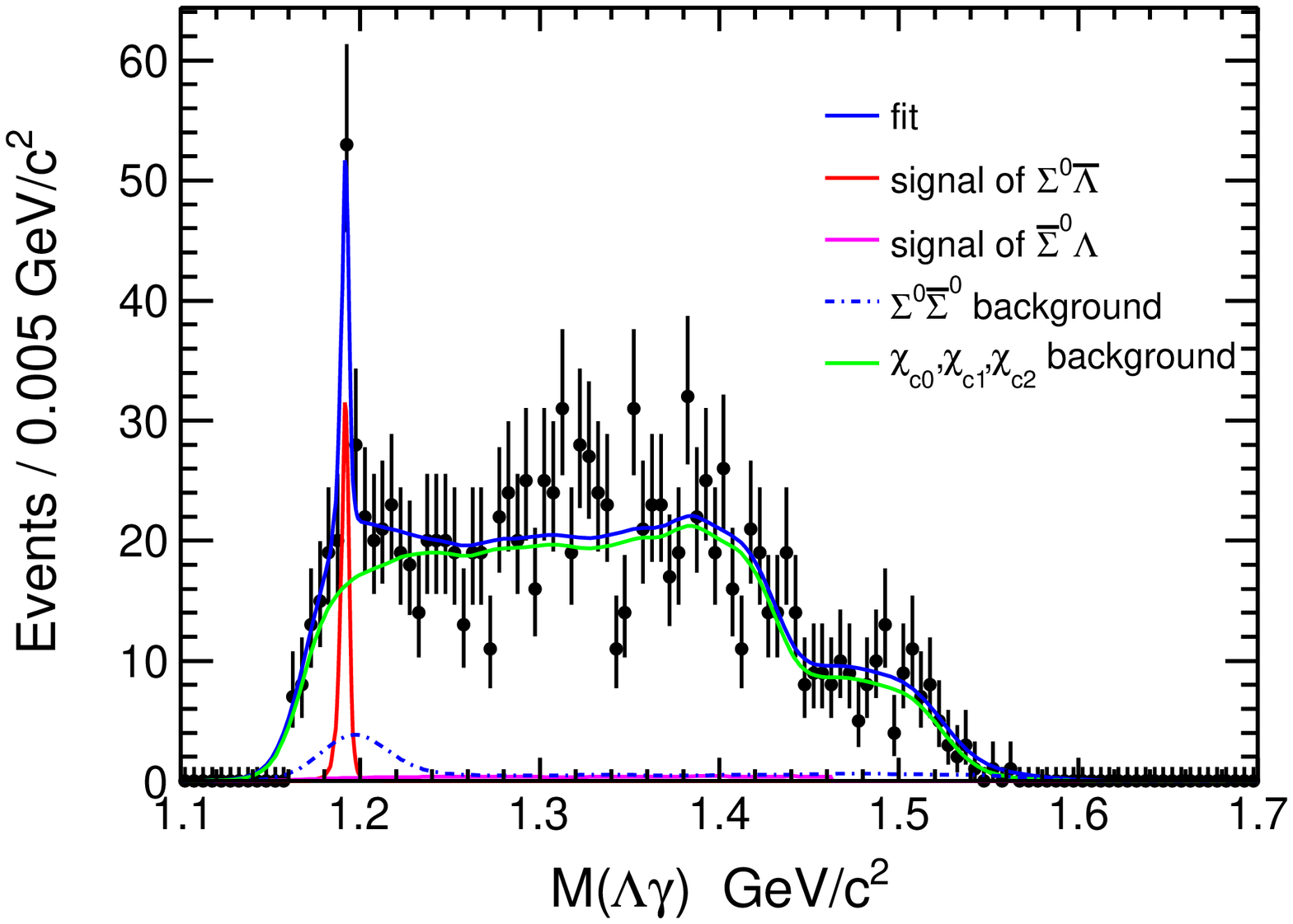}}\put(-200,150){\bf  ~(a)}
  \subfigure{\includegraphics[width=0.48\textwidth]{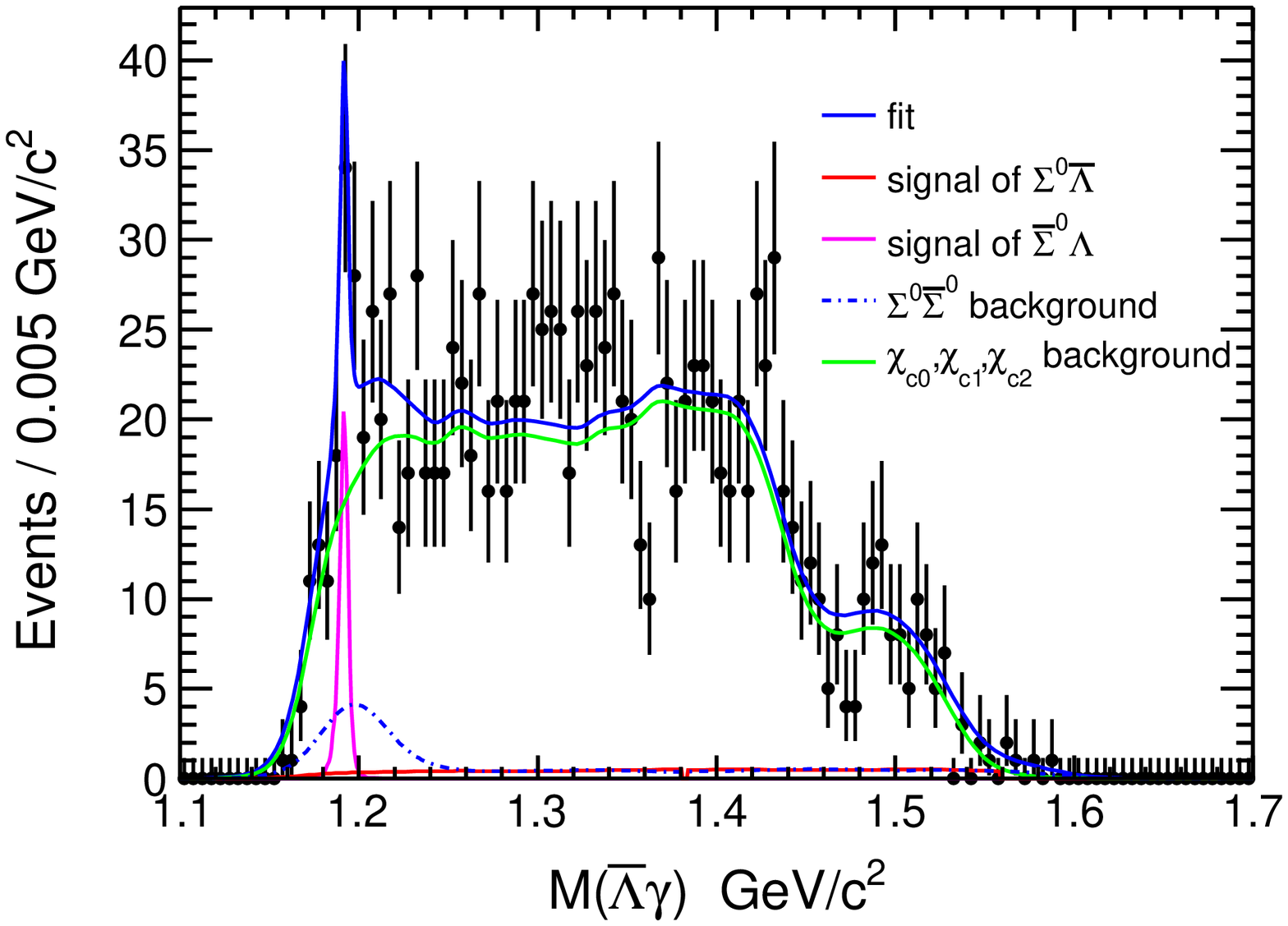}}\put(-200,150){\bf  ~(b)}
  \caption{The projections from the two-dimensional fit to $\rm
    M_{\gamma\Lambda}$ and $\rm M_{\gamma\bar{\Lambda}}$. Dots with
    error bars are data, blue solid curves are fitting results, red
    and pink curves are the signals, blue dotted lines are from
    normalized $\psi(3686) \rightarrow \bar{\Sigma}^{0} \Sigma^{0}$
    background contributions, green lines show the $\psi(3686)
    \rightarrow \gamma\chi_{cJ}$ background contributions, the contributions from other backgrounds are too small to be drawn on the plots.}
 \label{fitting}
\end{figure*}

 We determine the signal yields by an unbinned maximum likelihood fit
 to the two-dimensional distributions of the $\gamma\Lambda$ and
 $\gamma\bar{\Lambda}$ invariant masses. The signal shapes are
 determined from signal MC simulation for the $\Sigma^{0}
 \bar{\Lambda}$ and $\bar{\Sigma}^{0} \Lambda$ processes. The
 background shape includes five items: $\psi(3686) \rightarrow
 \Sigma^{0} \bar{\Sigma}^{0}$, $\psi(3686)\rightarrow \gamma
 \chi_{cJ}(\chi_{cJ} \rightarrow \Lambda \bar{\Lambda})$ with $J$ = 0,
 1, 2, and other background contributions. The shapes of the first
 four items are determined from MC simulation while the last one is
 described by a two-variable first-order polynomial function
 $f(m_{\gamma\Lambda}, m_{\gamma\bar{\Lambda}}) = am_{\gamma\Lambda}+
 bm_{\gamma\bar{\Lambda}}+c$ where $a$, $b$, $c$ are constant
 parameters that are determined in the fit. The background yields are
 floated in the fit except the peaking background $\psi(3686) \to
 \Sigma^{0} \bar{\Sigma}^{0}$ which is included with its magnitude
 determined from previous measurements~\cite{a2}.  The $\chi_{cJ}$
 background yields are consistent with expectation after considering
 branching fractions~\cite{PDG} and efficiencies.
 Figure~\ref{fitting} shows the projections of the two-dimensional
 fitting results. The numbers of signal events are determined to be
 $N^{\textrm{sig}}_{\gamma\bar{\Lambda}}= 26.1\pm6.6$,
 $N^{\textrm{sig}}_{\gamma\Lambda}= 37.2\pm7.7$ from the fit.

The contribution from the continuum process, $i.e.$
$e^{+}e^{-}\rightarrow \gamma^{*}\rightarrow \bar{\Sigma}^{0} \Lambda
+c.c.$ is estimated from the collision data at 3.773 GeV with
integrated luminosity of 2931.8 pb$^{-1}$ taken during 2010 and
2011. The same event selection criteria as for the $\psi(3686)
\rightarrow \bar{\Sigma}^{0} \Lambda +c.c. $ decay is
applied.
In addition, $|\rm M_{\Lambda \bar{\Lambda}}-3.686$
GeV/$c^{2}|>0.01$ GeV/$c^{2}$ is required to suppress background from the $e^{+} e^{-}
\rightarrow \gamma_{ISR}$ $\psi(3686)$, $\psi(3686) \rightarrow
\Lambda \bar{\Lambda}$ process.  An unbinned one dimensional maximum likelihood fit is
done to determine signal yields, where the peaking background $e^+e^-\rightarrow\bar{\Sigma}^{0}\Sigma^{0}$ has been considered with its shape from MC simulation and magnitude from previous measurements~\cite{a2}. The other backgrounds are described with a second order
polynomial function. To account for the difference of the
integrated luminosity and cross sections between the two energy points
3.686 GeV and 3.773 GeV, a scaling factor $f=0.24$ is applied.  The
continuum contributions at 3.686 GeV are determined to be: $N_{\gamma
  \bar{\Lambda}}^{\rm{cont}}=6.2~\pm~1.2$ and $N_{\gamma
  \Lambda}^{\rm{cont}}=3.9~\pm~1.0$ in the $\Sigma^{0} \bar{\Lambda}$ and
$\bar{\Sigma}^{0} \Lambda$ processes, respectively, where the
contributions from
$e^+e^-\rightarrow\psi(3770)\rightarrow\bar{\Sigma}^{0} \Lambda +c.c.$
decay have been ignored due to its low production.

The branching fraction of $\psi(3686)\rightarrow\bar{\Sigma}^{0}\Lambda$ is calculated by
\begin{equation}
\B\left(\psi(3686) \rightarrow \bar{\Sigma}^{0} \Lambda\right)=\frac{N^{\textrm{sig}}_{\gamma\bar{\Lambda}}-N_{\gamma\bar{\Lambda}}^{\textrm{cont}}}{N_{\psi(3686)} \cdot \epsilon_{\bar{\Sigma}^{0} \Lambda} \cdot Br }.
\end{equation}
Here, $N_{\psi(3686)}$ is the total number of $\psi(3686)$
events~\cite{totalnumber}, $Br~=~\B\left(\bar{\Lambda} \rightarrow
\bar{p} \pi^{+}\right) \cdot \B\left(\Lambda \rightarrow p
\pi^{-}\right) \cdot \B\left(\bar{\Sigma}^{0} \rightarrow \gamma
\bar{\Lambda}\right)$~\cite{PDG}, and the efficiency
$\epsilon_{\bar{\Sigma}^{0}\Lambda}= 16.52\%$ is determined from
simulation. The branching fraction
of~$\psi(3686)\rightarrow\bar{\Sigma}^{0}\Lambda$~is determined to be
$(0.66\pm0.22)\times10^{-6}$. Similarly, the branching fraction of
$\psi(3686)\rightarrow \Sigma^{0}\bar{\Lambda}$ is calculated to be
$(0.94\pm 0.22)\times 10^{-6}$, with $\epsilon_{\Sigma^{0}
  \bar{\Lambda}}= 19.44\%$. The clear difference between $\epsilon_{\Sigma^{0}
  \bar{\Lambda}}$ and $\epsilon_{\bar{\Sigma}^{0}\Lambda}$ comes from the different selection criteria on the open angle between photon and (anti-)proton. The combined branching fraction of
$\psi(3686)\rightarrow\bar{\Sigma}^{0}\Lambda+c.c.$ is
$\B\left(\psi(3686)\!\rightarrow\!\bar{\Sigma}^{0}\Lambda\!+\!c.c.\right)\!=(1.60\pm
0.31)\times 10^{-6}$, where the uncertainty is statistical only.

\subsection{Systematic uncertainties}
The systematic uncertainties on the branching-fraction measurement
include those from track and photon reconstruction efficiencies, kinematic
fit, angle requirement, $\Lambda(\bar{\Lambda})$ reconstruction efficiency, signal and background
shapes, and the branching fraction of $\Lambda(\bar{\Lambda})$ decay.

 The uncertainty due to photon detection efficiency is $1 \%$ per
 photon, which is determined from a study of the control sample $J /
 \psi \rightarrow \rho\pi$~\cite{photonEfficiency}.

 The efficiency of $\Lambda(\bar{\Lambda})$ reconstruction is studied
 using the control sample of $\psi(3686) \rightarrow$ $\Lambda
 \bar{\Lambda}$ decays, and a correction factor $0.980 \pm
 0.011$~\cite{trackEfficiency} is applied to the efficiencies obtained
 from MC simulation.  The uncertainty of the correction factor, 1.1\%
 already includes the uncertainties of MDC tracking and
 $\Lambda(\bar{\Lambda})$ reconstruction, and 1.1\% is taken as the
 uncertainty of the efficiency of $\Lambda(\bar{\Lambda})$
 reconstruction.

To study the uncertainty caused by the requirement on the angle
between the position of each shower in the EMC and the closest
extrapolated charged track, we utilize the processes
$\psi(3686)\rightarrow\gamma\chi_{cJ},\chi_{cJ}\rightarrow\Lambda
\bar{\Lambda}$ due to their large statistics. Two sets of branching
fractions are obtained. One is with the nominal requirement, {\it i.e.}
the angle of each shower is at least $10^{\circ}$ away from $p,
\pi^{+}, \pi^{-}$ tracks and $20^{\circ}$ away from $\bar{p}$ track;
the other one requires the angle of each shower is at least
$20^{\circ}$ away from $p, \pi^{+}, \pi^{-}$ tracks and $30^{\circ}$
away from $\bar{p}$ track. The difference of the branching fractions obtained with the nominal and
modified requirements is $1.6\%$, which is taken as the associated
systematic uncertainty.

To study the uncertainty associated with the kinematic fit, the track
helix parameters are corrected in the MC
simulation~\cite{kfitUncer}. The resulting $0.3 \%$ efficiency
difference before and after the correction is taken as the systematic
uncertainty related to the kinematic fit.

The systematic uncertainty associated with the signal shape is mainly
due to the resolution difference between data and MC simulation. It is
estimated by smearing the signal shape with a resolution of $5\%$ of the one determined from MC simulation according to the study using the control sample of
$\psi(3686)\rightarrow\bar{\Sigma}^{0}\Sigma^{0}$. The difference
is 0.4\% and is taken as the corresponding
systematic uncertainty due to the signal shape.

For the peaking background, we fixed the shape and number of events in
the fitting. We vary the number of background events by its uncertainty. A difference of 1.9\% is assigned as systematic uncertainty.

The uncertainty associated with the $\chi_{c0}$, $\chi_{c1}$, and $\chi_{c2}$
backgrounds is estimated by fixing their contributions to the
world average values~\cite{PDG} instead of floating them in the
nominal fit. The differences between fixing and floating are
0.7\%, 0.4\%, and 2.1\% for $\chi_{c0}$, $\chi_{c1}$, and $\chi_{c2}$,
respectively, and are taken as the systematic uncertainties.

The systematic uncertainty for the description of the other background
contributions is estimated by changing from a two-variable first-order
polynomial to a two-variable second-order polynomial to describe the
shape of the other backgrounds. The systematic uncertainty is determined to be 0.1\%.

In the signal MC sample, $\psi(3686) \rightarrow \bar{\Sigma}^{0}
\Lambda +c . c .$ is simulated with an uniform distribution over phase space. However, the angular distribution should be described as
$\mathrm{d}N / \mathrm{d} \cos \theta \propto 1+\alpha
\cos^{2}\theta$~\cite{physicsmode} where $\theta$ is the polar angle
of the (anti)~baryon. Since the observed number of events does not
allow the determination of the angular distribution in our analysis,
we generate MC samples with $\alpha=-1.0$ and $\alpha=1.0$, the two
extreme scenarios.  The efficiency difference between them is divided by
$\sqrt{12}$ under the assumption that the prior distribution of
$\alpha$ is uniform, and the result $6.9 \%$ is taken as the
uncertainty associated to the angular distribution.

 The uncertainty on the total number of $\psi(3686)$ events is $0.6
 \%$~\cite{totalnumber}.  The uncertainty of the $\Lambda$ decay branching
 fraction is taken from the world average
 value~\cite{PDG}.  Table~\ref{totaluncer}~lists all sources and
 values of systematic uncertainties, and the total systematic
 uncertainty is determined by adding them in quadrature.
\begin{table}[h]\small 
  \centering
\caption{Systematic uncertainties for the branching fraction of $\psi(3686)
  \rightarrow \bar{\Sigma}^{0} \Lambda+c.c.$ decay.}
\begin{tabular}{lc}
\hline \multicolumn{1}{c} { Source } & $\psi(3686) \rightarrow \bar{\Sigma}^{0} \Lambda +$ c.c. $(\%)$ \\
\hline Photon efficiency & 1.0 \\
$\Lambda$ efficiency correction &  1.1 \\
Angle requirement &  1.6 \\
Kinematic fit & 0.3 \\
Signal shape & 0.4 \\
Peaking background & 1.9 \\
Background of $\psi(3686)\to \gamma \chi_{c0}$ & 0.7 \\
Background of $\psi(3686)\to \gamma\chi_{c1}$ & 0.4 \\
Background of $\psi(3686)\to \gamma\chi_{c2}$ & 2.1 \\
Other non-resonance background & 0.1\\
Physics model &  6.9 \\
$\B\left(\Lambda \rightarrow p \pi\right)$ & 1.1 \\
Number of $\psi(3686)$ & 0.6 \\
\hline Total &   7.9 \\
\hline
\end{tabular}
  \label{totaluncer}
\end{table}

\subsection{Discussion and result}
So far, we have not considered possible interference between the
$\psi(3686)$ decay and the continuum process, which is
described by \begin{equation}\left|A_{\textrm{cont}} + e^{i\theta}
  A_{\psi^{\prime}}\right|^{2}\propto
  N^{\textrm{sig}}_{\gamma\bar{\Lambda}}(N^{\textrm{sig}}_{\gamma\Lambda}),
  \end{equation} where
$A_{\psi^{\prime}}$~and $A_{\textrm{cont}}$ are the amplitudes of
$\psi(3686)\rightarrow\bar{\Sigma}^{0}\Lambda+c.c.$ and the continuum
contribution, respectively.
The difference between $\theta=0^{\circ}$ and
$\theta=180^{\circ}$, corresponding to the extreme constructive and
destructive cases respectively, is adopted as the uncertainty
associated with the interference. This difference is divided by
$\sqrt{12}$, since the prior distribution of the interference angle is
assumed to be uniform. Finally, the branching fractions are $\B\left(\psi(3686)
\rightarrow \Sigma^{0}\bar{\Lambda}\right)=(0.94~\pm~0.39) \times
10^{-6}$ and $\B\left(\psi(3686) \rightarrow
\bar{\Sigma}^{0}\Lambda\right)=(0.66~\pm~0.49) \times 10^{-6}$, where the uncertainties are only the systematic arising from interference.  The
combined branching fraction of $\psi(3686) \rightarrow
\bar{\Sigma}^{0} \Lambda +c.c.$ is $\B\left(\psi(3686) \rightarrow
\bar{\Sigma}^{0} \Lambda+c.c.\right)=(1.60~\pm~0.31~\pm~0.13~\pm~0.58)
\times 10^{-6}$, where the first uncertainty is statistic, the second
is systematic, and the third is the uncertainty due to interference with
the continuum.

\section{\boldmath $\chi_{cJ}\rightarrow\Lambda\bar{\Lambda}$}
\subsection{Event selection and background study}
The initial selection criteria for charged tracks and photons and
the $\Lambda(\bar{\Lambda})$ reconstruction are the same as those
described above for $\psi(3686) \rightarrow \bar{\Sigma}^{0} \Lambda
+c.c.$.
Additional selection criteria are 1.) the $\chi^{2}$ from the $6 \mathrm{C}$
kinematic fit is required to be less than 50, and 2.)  to veto
$\psi(3686) \rightarrow \Sigma^{0} \bar{\Sigma}^{0}$ background, the
$\gamma \Lambda(\gamma \bar{\Lambda})$ combination is required to be
outside of the $\Sigma^{0}\left(\bar{\Sigma}^{0}\right)$ region, that
is defined as within $12$ MeV to the $\Sigma^0$ nominal mass~\cite{PDG}.
The $\chi_{c0}$ veto is also removed.

\begin{figure}[htb]
  \centering
  \includegraphics[width=8cm]{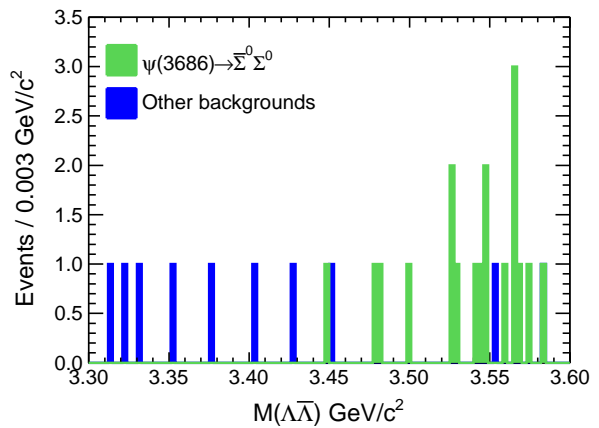}\\
  \caption{The $\chi_{cJ} \rightarrow \Lambda
  \bar{\Lambda}$ background distributions from the inclusive MC
  samples. The green histogram is the $\Sigma^{0}\bar{\Sigma}^{0}$
  background, and the blue one is all the other background
  contributions.}
  \label{chicjback}
\end{figure}

After all selection requirements have been applied,
Fig.~\ref{chicjback} shows all background contributions according to
the inclusive MC sample, which include two parts: non-flat $\Sigma^{0}
\bar{\Sigma}^{0}$ background contributions that tend to accumulate in
the $\chi_{c 2}$ region and flat non- $\Sigma^{0}$ background
contributions. Both background levels are quite low compared with the
signal.


\subsection{Signal yields and branching fractions}
\begin{figure}[htb]
  \centering
  \includegraphics[width=8cm]{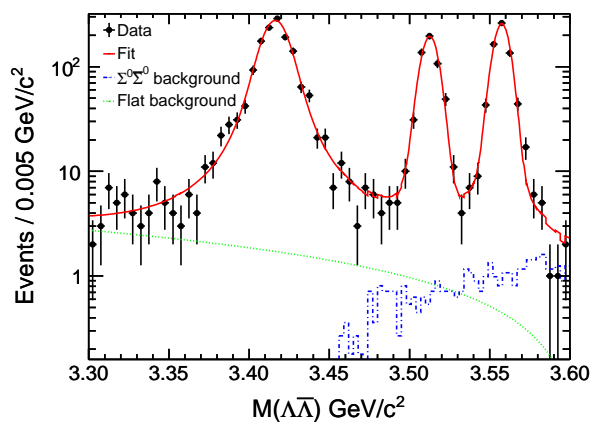}\\
  \caption{Fitting results of the $\Lambda \bar{\Lambda}$ invariant mass
    distribution. Dots with error bars are data, the red solid line is
    the fitting curve, the blue dashed line is $\Sigma^{0}
    \bar{\Sigma}^{0}$ background and the green dashed line is all
    other background contributions.}
  \label{fitchicj}
\end{figure}

To determine the $\chi_{cJ}$ signal yields, we fit the $\Lambda
\bar{\Lambda}$ invariant mass distribution with an
unbinned maximum likelihood fit. Each $\chi_{cJ}$ signal shape is
described by a Breit-Wigner function convolved with a Gaussian
function, and the parameters of the Breit-Wigner functions are fixed
to the world average values~\cite{PDG}. The Gaussian function
represents the resolutions, whose parameters are floated in the fit
but shared with all three $\chi_{cJ}$ resonances. The background shape
is composed of two parts: a MC simulation of $\psi(3686) \rightarrow
\Sigma^{0} \bar{\Sigma}^{0}$ events with both shape and number fixed
and a second-order polynomial with floating parameters to describe
other background contributions. The results are shown in
Fig.~\ref{fitchicj}. The amount of other backgrounds from the fit is obvious larger than that simulated in inclusive MC as shown in Fig.~\ref{chicjback}. It indicates the inclusive MC simulation is not perfect yet. The fitted $\chi_{c J}$ signal yields
 are $N_{\chi_{c 0}}=1486 \pm 42$, $N_{\chi_{c 1}}=528 \pm
24$, $N_{\chi_{c 2}}=670 \pm 27$.

The product branching fractions of $\B\left(\chi_{c J} \rightarrow
\Lambda \bar{\Lambda}\right) \cdot \B\left(\psi(3686) \rightarrow
\gamma \chi_{c J}\right)$ are determined by
\begin{equation}
\begin{aligned}
&\mathcal{B}\left(\chi_{c J} \rightarrow \Lambda \bar{\Lambda}\right) \cdot \mathcal{B}\left(\psi(3686) \rightarrow \gamma \chi_{c J}\right)=\\
&\frac{N_{\chi_{cJ}}}{N_{\psi(3686)} \cdot \epsilon \cdot \mathcal{B}\left(\Lambda \rightarrow p \pi^{-}\right) \cdot \mathcal{B}\left(\bar{\Lambda} \rightarrow \bar{p} \pi^{+}\right)},
\end{aligned}
\end{equation}
 where $\epsilon$ is the efficiency. We calculate the branching fractions of $\chi_{c J}
 \rightarrow \Lambda \bar{\Lambda}$ decays based on world averaged
 values of $\mathcal{B}\left(\psi(3686) \rightarrow \gamma
 \chi_{c J}\right)$~\cite{PDG}. The results are listed in
 Table~\ref{fitingResulte2}.

\subsection{ Systematic uncertainties}

The uncertainties in the branching-fraction measurement include
photon and $\Lambda(\bar{\Lambda})$ reconstruction efficiencies, kinematic fit, signal and
background shapes, fitting range, and the branching fraction of $\Lambda(\bar{\Lambda})$ decay.

The uncertainties due to the photon detection and $\bar{\Lambda}$
reconstruction efficiencies, the requirement on the angle between the
position of each shower in the EMC and the closest extrapolated
charged tracks, the $\Lambda$ branching fraction, and the number of
$\psi(3686)$ events are the same as in the study of $\psi(3686)
\rightarrow \bar{\Sigma}^{0} \Lambda +c.c.$, while that due to the
kinematic fit is calculated in the same manner.

For the signal shapes,
the single Gaussian is changed to a double Gaussian, and the
differences in the yields of signal events, $0.6 \%, 0.3 \%,$ and $0.1
\%$ for $\chi_{c 0}, \chi_{c 1},$ and $\chi_{c 2},$ respectively, are
taken as the systematic uncertainties associated with the signal
shapes. We study the uncertainty associated with other backgrounds shape by changing description from the second-order polynomial function to the third-order polynomial function. It turns out the difference is negligible.
 \begin{table}[htb]\normalsize \centering
 \renewcommand\arraystretch{1.2} \vspace{0.2cm} \noindent
 \caption{Systematic uncertainties of the branching fractions for $\chi_{c
 J} \rightarrow \Lambda \bar{\Lambda}$ decays.}  \begin{tabular}{lccc}
 \hline \multicolumn{1}{c} { Source } & $\chi_{c 0}(\%)$ & $\chi_{c
 1}(\%)$ & $\chi_{c 2}(\%)$ \\ \hline Photon efficiency & 1.0 & 1.0 &
 1.0 \\ $\Lambda$ efficiency correction & 1.1 & 1.1 & 1.1 \\ Kinematic fit & 0.5
 & 0.8 & 0.8 \\ Angle requirement & 1.6 & 1.6 & 1.6 \\ Signal shape & 0.6 &
 0.3 & 0.1 \\ Peaking background & 0.0 & 0.2 & 0.4 \\ Fitting range &
 0.6 & 1.8 & 0.7 \\ Angular distribution & 0.0 & 3.0 & 8.7 \\
 $\B\left(\Lambda \rightarrow p \pi\right)$ & 1.1 & 1.1 & 1.1 \\
 Number of $\psi(3686)$ & 0.6 & 0.6 & 0.6 \\ \hline sum & 2.7 & 4.4 &
 9.1 \\ $\B\left(\psi(3686) \rightarrow \gamma \chi_{c J}\right)$ &
 2.0 & 2.5 & 2.1 \\ \hline Total & 3.4 & 5.1 & 9.4 \\ \hline
 \end{tabular} \label{chicjuncer} \end{table}

 The uncertainty associated with the fitting range is estimated by
 varying it from [3.30, 3.60] $\mathrm{GeV}$ to [3.32, 3.62]
 $\mathrm{GeV}.$ The differences 0.6\%, 1.8\%, and 0.7\% for $\chi_{c 0}$,
 $\chi_{c 1}$, and $\chi_{c 2}$ are taken as the uncertainty due to the
 fitting range.

\begin{table*}
 \renewcommand{\arraystretch}{1.3}
  \centering
  \noindent
  \normalsize
  \caption{ The number of observed events $N_{\chi_{cJ}}$, efficiencies ($\epsilon$), product branching fractions, and the branching fractions of $\chi_{c J} \rightarrow \Lambda
    \bar{\Lambda}$ decays compared with the world average values,
    where the
    third uncertainties for $\B\left(\chi_{c J} \rightarrow \Lambda
    \bar{\Lambda}\right)$ are the uncertainties due to the branching
    fractions of $\psi(3686) \rightarrow \gamma \chi_{c J}$ decays.}
\begin{tabular}{c c c p{5cm}<{\centering} p{4cm}<{\centering}p{3cm}<{\centering}}
\hline  \multirow{2}{*} {Mode} & \multirow{2}{*}{$N_{\chi_{cJ}}$} & \multirow{2}{*}{$\epsilon$} & $\B\left(\psi(3686) \rightarrow \gamma \chi_{c J}\right)$ & \multicolumn{2}{c} {$\B\left(\chi_{c J} \rightarrow \Lambda \bar{\Lambda}\right)\left(\times 10^{-4}\right)$} \\
\cline { 5 - 6 } & & & $\times \B\left(\chi_{c J}
\rightarrow \Lambda \bar{\Lambda}\right)\left(10^{-5}\right)$ & This work & $\mathrm{PDG}$ \\
\hline$\chi_{c 0}$ & $1486 \pm 42$ & 22.80\%&  $ 3.56 \pm 0.10 \pm  0.10$ & $ 3.64 \pm 0.10 \pm  0.10 \pm 0.07$ & $3.27 \pm 0.24$ \\
$\chi_ {c 1}$ & $528 \pm 24$ & 22.61\%&  $ 1.28 \pm 0.06 \pm 0.06$ & $  1.31 \pm 0.06 \pm  0.06\pm 0.03$ & $1.14 \pm 0.11$ \\
$\chi_{c 2}$ & $670 \pm 27$ & 20.16\% & $ 1.82 \pm 0.08 \pm 0.17$ & $ 1.91 \pm 0.08 \pm   0.17 \pm 0.04$ & $1.84 \pm 0.15$ \\
\hline
\end{tabular}
\label{fitingResulte2}
\end{table*}

The angular distributions of $\psi(3686) \rightarrow \gamma \chi_{c
  1,2}$ are known. However, knowledge of the $\chi_{c 1,2} \rightarrow
  \Lambda \bar{\Lambda}$ angular distributions is still limited. We
  generate signal MC samples with an uniform distribution over phase space. To estimate the
  uncertainty caused by the angular distribution, we adopt the method
  used in Ref.~\cite{Beschicj}. We regenerate the signal MC samples
  according to the helicity amplitudes $B_{\lambda_3,\bar{\lambda}_3}$ defined in Ref.~\cite{angleUncer}, where $\lambda_3 (\bar{\lambda}_3)$ is the helicity of $\Lambda (\bar{\Lambda})$ in the rest frame of $\chi_{cJ}$.
The amplitudes  $B_{\frac{1}{2}, \frac{1}{2}}$ and $B_{\frac{1}{2}, -\frac{1}{2}}$ are both set to be $1.0$ to obtain the efficiency again, and
  the differences of $ 3.0 \%$ and $ 8.7 \%$ between these two models
  are taken as the associated systematic uncertainties.

Table~\ref{chicjuncer}~lists all sources of systematic uncertainty and
the values for each decay channel. The total systematic uncertainties are
determined by adding each contribution in quadrature.

\subsection{Result} The branching fractions, compared with the world
averaged values, are listed in {Table \ref{fitingResulte2}}, where the
third uncertainties for $\mathcal{B}\left(\chi_{c J} \rightarrow \Lambda
\bar{\Lambda}\right)$ are the uncertainties due to the branching fractions
of $\psi(3686) \rightarrow \gamma \chi_{c J}$.

\section{summary}
 The branching fraction of the isospin symmetry breaking decay
 $\psi(3686)\rightarrow\bar{\Sigma}^{0}\Lambda+c.c.$ is measured to be
 $\B\left(\psi(3686) \rightarrow \bar{\Sigma}^{0}\Lambda +
 c.c.\right)=(1.60\pm 0.31 \pm 0.13 \pm 0.58) \times 10^{-6}$, where
 the first uncertainty is statistical, the second is systematic, the
 third one is the uncertainty due to interference with the
 continuum. Compared with the result using CLEO-c data~\cite{a2}, $(12.3 \pm 2.4)
 \times 10^{-6}$, our result is significantly smaller. Our measurement
 is consistent with the theoretical prediction~\cite{a4}, $(4.0 \pm
 2.3)\times10^{-6}$, within 1$\sigma$.  However our branching fraction
 is measured under the assumption of no interference which corresponds
 to an angle of $\theta = 90^{\circ}$, while the angle is assumed to
 be $0^{\circ}$ in Ref.~\cite{a4}.  If $\theta=0^{\circ}$ is adopted,
 the branching fraction is measured to be $\B\left(\psi(3686)
 \rightarrow \bar{\Sigma}^{0}\Lambda + c.c.\right)=(1.02\pm 0.31 \pm
 0.13) \times 10^{-6}$, and the difference between our measurement and
 the theoretical prediction is larger than 1$\sigma$ but still smaller
 than 2$\sigma$.

With the increased data sample collected at the BESIII detector, the
branching fractions of $\chi_{cJ}\to \Lambda\bar{\Lambda} $ are
measured via $\psi(3686)\rightarrow\gamma\chi_{cJ}$ with improved
precision. The branching fractions are determined to be
$\B\left(\chi_{c 0} \to \Lambda \bar{\Lambda}\right)=(3.64 \pm 0.10
\pm 0.10 \pm 0.07) \times 10^{-4}$, $\B\left(\chi_{c1}
\rightarrow\Lambda\bar{\Lambda}\right)=(1.31 \pm 0.06 \pm 0.06 \pm
0.03) \times 10^{-4}$, $\B\left(\chi_{c 2} \rightarrow \Lambda
\bar{\Lambda}\right)=(1.91 \pm 0.08 \pm 0.17 \pm 0.04) \times
10^{-4}$, where the first and second uncertainties are statistical and
systematic, and the third ones are the systematic uncertainties due to the
uncertainties on the $\psi(3686) \to \gamma \chi_{cJ}$ branching
fractions. These results, which supersede the previous BESIII
measurements of branching fractions ($\chi_{cJ}\to
\Lambda\bar{\Lambda}$) in Ref.~\cite{Beschicj}, are consistent with
the world average values \cite{PDG}, but not with the theoretical
predictions~\cite{36,37,38}. This should be understood.

\begin{acknowledgments}
The BESIII collaboration thanks the staff of BEPCII and the IHEP computing center for their strong support. This work is supported in part by National Key Research and Development Program of China under Contracts Nos. 2020YFA0406300, 2020YFA0406400; National Natural Science Foundation of China (NSFC) under Contracts Nos. 11875115, 11625523, 11635010, 11735014, 11822506, 11835012, 11935015, 11935016, 11935018, 11961141012; the Chinese Academy of Sciences (CAS) Large-Scale Scientific Facility Program; Joint Large-Scale Scientific Facility Funds of the NSFC and CAS under Contracts Nos. U1732263, U1832207,U2032110; CAS Key Research Program of Frontier Sciences under Contracts Nos. QYZDJ-SSW-SLH003, QYZDJ-SSW-SLH040; 100 Talents Program of CAS; INPAC and Shanghai Key Laboratory for Particle Physics and Cosmology; ERC under Contract No. 758462; European Union Horizon 2020 research and innovation programme under Contract No. Marie Sklodowska-Curie grant agreement No 894790; German Research Foundation DFG under Contracts Nos. 443159800, Collaborative Research Center CRC 1044, FOR 2359, FOR 2359, GRK 214; Istituto Nazionale di Fisica Nucleare, Italy; Ministry of Development of Turkey under Contract No. DPT2006K-120470; National Science and Technology fund; Olle Engkvist Foundation under Contract No. 200-0605; STFC (United Kingdom); The Knut and Alice Wallenberg Foundation (Sweden) under Contract No. 2016.0157; The Royal Society, UK under Contracts Nos. DH140054, DH160214; The Swedish Research Council; U. S. Department of Energy under Contracts Nos. DE-FG02-05ER41374, DE-SC-0012069.
\end{acknowledgments}

\bibliography{draft}

\begin{thebibliography}{99}


\bibitem{a3}
D. M. Asner $et~al$.,
Int. J. Mod. Phys. A {\bf 24}, 499 (2009).
\bibitem{polarization}
M. Ablikim {\it et al}., [BESIII Collaboration], Nature Phys. {\bf 15}, 631-634 (2019).

\bibitem{CLEO-measurement}
 T. K. Pedlar {\it et al}., [CLEO Collaboration], Phys. Rev. D {\bf 72}, 051108 (2005).
\bibitem{BES-measurement1}
M. Ablikim {\it et al}., [BESIII Collaboration], Phys. Lett. B {\bf 648}, 149 (2007).
\bibitem{BES-measurement2}
M. Ablikim {\it et al}., [BESIII Collaboration], Phys. Lett. B {\bf 632}, 181 (2006).
\bibitem{BES-measurement3}
M. Ablikim {\it et al}., [BESIII Collaboration], Phys. Rev. D {\bf 78}, 092005 (2008).
\bibitem{BESIII-measurement1}
M. Ablikim {\it et al}., [BESIII Collaboration], Phys. Rev. D {\bf 86}, 032014 (2012).
\bibitem{BESIII-measurement2}
M. Ablikim {\it et al}., [BESIII Collaboration], Phys. Rev. D {\bf 86}, 032008 (2012).
\bibitem{BESIII-measurement3}
M. Ablikim {\it et al}., [BESIII Collaboration], Phys. Rev. D {\bf 93}, 072003 (2016).

\bibitem{1803.02039}
M.~Ablikim \textit{et al.,} [BESIII Collaboration],
Phys. Rev. D \textbf{98}, 032006 (2018).

\bibitem{1907.13041}
M.~Ablikim \textit{et al.,} [BESIII Collaboration],
Phys. Rev. D \textbf{100}, 051101 (2019).

\bibitem{1911.06669}
M.~Ablikim \textit{et al.,} [BESIII Collaboration],
Phys. Rev. D \textbf{101}, 012004 (2020).

\bibitem{2004.07701}
M.~Ablikim \textit{et al.,} [BESIII Collaboration],
Phys. Rev. Lett. \textbf{125}, 052004 (2020).

\bibitem{2007.03679}
M.~Ablikim \textit{et al.,} [BESIII Collaboration],
arXiv:2007.03679 [hep-ex].


\bibitem{a4}
K. Zhu, X. H. Mo, C. Z. Yuan, P. Wang, Int. J. Mod. Phys. A {\bf 30}, 1550148 (2015).


\bibitem{a2}
S. Dobbs, Kamal K. Seth, A. Tomaradze, T. Xiao, and G. Bonvicini, Phys. Rev. D {\bf 96}, 092004 (2017).
\bibitem{Ferroli:2020xnv}
R.~B.~Ferroli, A.~Mangoni and S.~Pacetti,
Eur. Phys. J. C \textbf{80}, 903 (2020)

\bibitem{Beschicj}
M. Ablikim {\it et al}., [BESIII Collaboration], Phys. Rev. D {\bf 87}, 032007 (2013).

 \bibitem{Ablikim:2009aa}
  M.~Ablikim {\it et al.}, [BESIII Collaboration],
  Nucl.\ Instrum.\ Meth.\ A {\bf 614}, 345 (2010).

\bibitem{Yu:IPAC2016-TUYA01}
   C.~H.~Yu {\it et al.},
  Proceedings of IPAC2016, Busan, Korea, (2016).


  \bibitem{Ablikim:2019hff}
  M.~Ablikim {\it et al.}, [BESIII Collaboration],
  Chin. Phys. C {\bf 44}, 040001 (2020).

\bibitem{geant4}
  S.~Agostinelli {\it et al.}, [{Geant4} Collaboration],
  Nucl.\ Instrum.\ Meth.\ A {\bf 506}, 250 (2003).

\bibitem{ref:kkmc}
  S.~Jadach, B.~F.~L.~Ward and Z.~Was,
  Phys.\ Rev.\ D {\bf 63}, 113009 (2001);
  Comput.\ Phys.\ Commun.\  {\bf 130}, 260 (2000).

\bibitem{ref:evtgen}
  D.~J.~Lange,
  Nucl.\ Instrum.\ Meth.\ A {\bf 462}, 152 (2001);
  R.~G.~Ping,
  Chin. Phys. C {\bf 32}, 599 (2008).
\bibitem{PDG}
P.~A.~Zyla \textit{et al.}, [Particle Data Group],
Prog. Theor. Exp. Phys. \textbf{2020}, 083C01 (2020).

\bibitem{ref:lundcharm}
  J.~C.~Chen, G.~S.~Huang, X.~R.~Qi, D.~H.~Zhang and Y.~S.~Zhu,
  Phys.\ Rev.\ D {\bf 62}, 034003 (2000);
  R.~L.~Yang, R.~G.~Ping and H.~Chen,
  Chin.\ Phys.\ Lett\  {\bf 31}, 061301 (2014).

\bibitem{photos}
  E.~Richter-Was,
  Phys.\ Lett.\ B {\bf 303}, 163 (1993).
\bibitem{angleUncer}
G. R. Liao, R. G. Ping, and Y. X. Yang, Chin. Phys. Lett {\bf 26}, 051101 (2009).
\bibitem{secondvertex}
 M. Xu {\it et al}., Chin. Phys. C {\bf 33}, 428 (2009).

\bibitem{totalnumber}
M. Ablikim $et~al$., [BESIII Collaboration],  Chin. Phys. C {\bf 42}, 023001 (2018).


\bibitem{photonEfficiency}
M. Ablikim {\it et al}., [BESIII Collaboration], Phys. Rev. D {\bf 83}, 112005 (2011).
\bibitem{trackEfficiency}
M. Ablikim {\it et al}., [BESIII Collaboration], Phys. Rev. D {\bf 95}, 052003 (2017).
\bibitem{kfitUncer}
M. Ablikim {\it et al}., [BESIII Collaboration], Phys. Rev. D {\bf 87}, 012002 (2013).
\bibitem{physicsmode}
P. Kessler, Nucl. Phys. B {\bf 15}, 253-266 (1970).




\bibitem{36}
R. G. Ping, B. S. Zou, and H. C. Chiang, Eur. Phys. J. A {\bf 23}, 129 (2005).
\bibitem{37}
X. H. Liu and Q. Zhao, J. Phys. G {\bf 38}, 035007 (2011).
\bibitem{38}
S. M. Wong, Eur. Phys. J.  C {\bf 14}, 643 (2000).
\end{thebibliography}

\end{document}